\documentclass[preprint,12pt]{elsarticle}
\usepackage{subfigure}
\usepackage{amsmath}
\usepackage{amssymb}
\usepackage{url}
\biboptions{sort&compress}
\usepackage[a4paper,total={6in, 8in}]{geometry}
\usepackage{atbegshi}
\AtBeginDocument{\AtBeginShipoutNext{\AtBeginShipoutDiscard}}

\journal{Journal of Magnetism and Magnetic Materials}

\DeclareMathOperator{\sign}{sign}

\begin{document}

\begin{frontmatter}

\title{Stability of skyrmion crystal phase in antiferromagnetic triangular lattice with DMI and single-ion anisotropy}
\author[1]{M. Mohylna},
\author[1]{M. \v{Z}ukovi\v{c}\corref{cor1}}
\ead{milan.zukovic@upjs.sk}
\address[1]{Department of Theoretical Physics and Astrophysics, Institute of Physics, Faculty of Science, Pavol Jozef \v{S}af\'arik University in Ko\v{s}ice, Park Angelinum 9, 041 54 Ko\v{s}ice, Slovak Republic}
\cortext[cor1]{Corresponding author}



\begin{abstract}

We study a frustrated antiferromagnetic Heisenberg model on a triangular lattice with the Dzyaloshinskii-Moriya interaction (DMI) in the presence of an external magnetic field and a single-ion anisotropy. Phase diagrams in the temperature-field plane for both easy-axis and easy-plane anisotropy of a varying strength are constructed in the regimes of a moderate and strong DMI by parallel tempering Monte Carlo simulations. For the considered range of parameters the phase diagrams featuring up to four ordered phases are identified. A skyrmion lattice (SkX) phase is found to be stabilized within some temperature and field window for sufficiently large DMI and not too strong anisotropy. For the systems with moderate (larger) DMI, a small easy-plane (easy-axis) anisotropy is concluded to be beneficial by extending (shifting) the field window of the SkX appearance to lower values. The Metropolis dynamics is employed to probe the persistence of SkX upon quenching the field to zero or small finite values.  The most favorable conditions for the skyrmions persistence are confirmed at low temperatures, small DMI and small easy-plane anisotropy values.

\end{abstract}



\end{frontmatter}

\section{Introduction}
Magnetic skyrmions are spin configurations of non-trivial topology currently actively investigated due to their unusual properties~\cite{nagaosa2013topological}, which make them suitable candidates for a number of different spintronic applications~\cite{sampaio2013nucleation, fert2013skyrmions, schulz2012emergent,zhang2015magnetic}. In recent years the interest shifted from the ferromagnets (FM), where the skyrmions were initially found~\cite{muhlbauer2009skyrmion}, to new potentially skyrmion-hosting materials. For instance, high expectations are put on antiferromagnetic (AFM) materials~\cite{kravchuk2019spin, gorobets20203d}, implementation of which in skyrmion-based devices is reported to have certain advantages over the FM ones~\cite{zhang2016antiferromagnetic, barker2016static}. Among other mechanisms, which can lead to the stabilization of a skyrmion phase in bulk samples and thin films~\cite{okubo2012multiple, heinze2011spontaneous,lin1973bubble}, is the inversion-symmetry-breaking Dzyaloshinskii-Moriya interaction, responsible for the creation of both N\'eel- and Bloch-type skyrmions of the size of 5-100 nm with fixed helicity. 

Although in theory skyrmions are considered to be stable against the thermal or quantum decay due to them being topologically protected, in reality they emerge as metastable structures with finite energy barriers preventing them from collapsing. Several works have studied the lifetime of an isolated skyrmion in thin films, multilayers and racetracks by means of the Monte Carlo (MC)~\cite{bessarab2018lifetime, rozsa2016complex} and the minimum-energy path~\cite{stosic2017paths, rohart2016path} methods. Mechanisms of the skyrmions' collapse in infinite samples have also been discussed~\cite{rohart2016path,lobanov2016mechanism,derras2019thermal}. Boundaries of the sample are found to play a crucial role in the destruction of a single skyrmion~\cite{bessarab2018lifetime}. A local decrease of the DMI in overall high DMI magnets has been proposed as a way to improve their thermal stability~\cite{stosic2017paths}. Relaxation dynamics of zero-field skyrmions to the stripes over a wide temperature range was described through three distinct processes for the skyrmions isolated, in the interior or at the boundary of a skyrmion lattice~\cite{peng2018relaxation}. The influence of the narrowness of the saddle point for Bloch- and N\'eel-type skyrmions and antiskyrmions on their stability has been shown to be of crucial importance~\cite{desplat2018thermal}. A huge role in the increase of the isolated skyrmions stability in a ferromagnetic system has been ascribed to the DMI~\cite{rohart2016path}. Frustration of the long-range interactions has been shown to enhance the skyrmion stability in systems with a ferromagnetic ground state~\cite{von2019enhanced}. Nevertheless, recently, Bessarab~\emph{et~al.}~\cite{bessarab2019stability} have demonstrated that in AFMs the skyrmion lattice stability can be enhanced by an application of an external magnetic field, which is in sharp contrast to the behaviour of their FM counterparts.

The idea of potential benefits of using frustration in the system to enhance the skyrmion stability is being actively developed. Okubo~\emph{et~al.}~\cite{okubo2012multiple} pointed out that both skyrmion and antiskyrmion lattices can be formed on any lattices of the trigonal symmetry with next-nearest interactions mainly due to the presence of the frustration even in the absence of the DMI. Isolated skyrmions and the skyrmion lattice are predicted to be stabilized in frustrated magnets with the DMI~\cite{mutter2019skyrmion} and single-ion anisotropy~\cite{leonov2015multiply}, respectively. Rosales~\emph{et~al.}~\cite{rosales2015three} have demonstrated that the combination of the frustration and the DMI in the AFM classical Heisenberg model on the triangular lattice leads to the stabilization of the skyrmion phase in a quite wide temperature-field window, which does not occur in an unfrustrated model. The study of a quantum model of such system suggested the existence of the skyrmion phase at relatively high temperatures~\cite{liu2020theoretical}. The connection between the skyrmions in the AFM triangular Heisenberg lattice with the DMI and $\mathbb{Z}_2$ vortices in the model without it was pointed out by Osorio~\emph{et~al.}~\cite{osorio2019skyrmions}. A very recent report by Fang~\emph{et~al.}~\cite{fang2021} suggests that such a skyrmion lattice can be realized in the presence of experimentally feasible external magnetic fields in interfaces integrating transition metal dichalcogenides with magnetic transition metals, such as Fe/MoS\textsubscript{2}.  

In our recent study we have considered the frustrated AFM triangular Heisenberg lattice with the DMI in an external magnetic field and studied the formation and evolution of the AFM skyrmion crystal in a wide parameter space~\cite{mohylna2021formation}. By a parallel tempering approach and a finite-size scaling analysis we found out that already very small DMI intensity can lead to the appearance of the skyrmion phase within some field window at sufficiently low temperatures. In the present paper we aim to study the influence of the single-ion anisotropy on the overall phase diagram topology with a particular focus on the skyrmion phase stability. Furthermore, we explore effects of different parameters on the skyrmion phase persistence (skyrmion lifetime) after quenching the external magnetic field to zero or small values. 


\section{Model and Method}

\begin{figure}[t!]
\centering
\subfigure{\includegraphics[scale=0.20,clip]{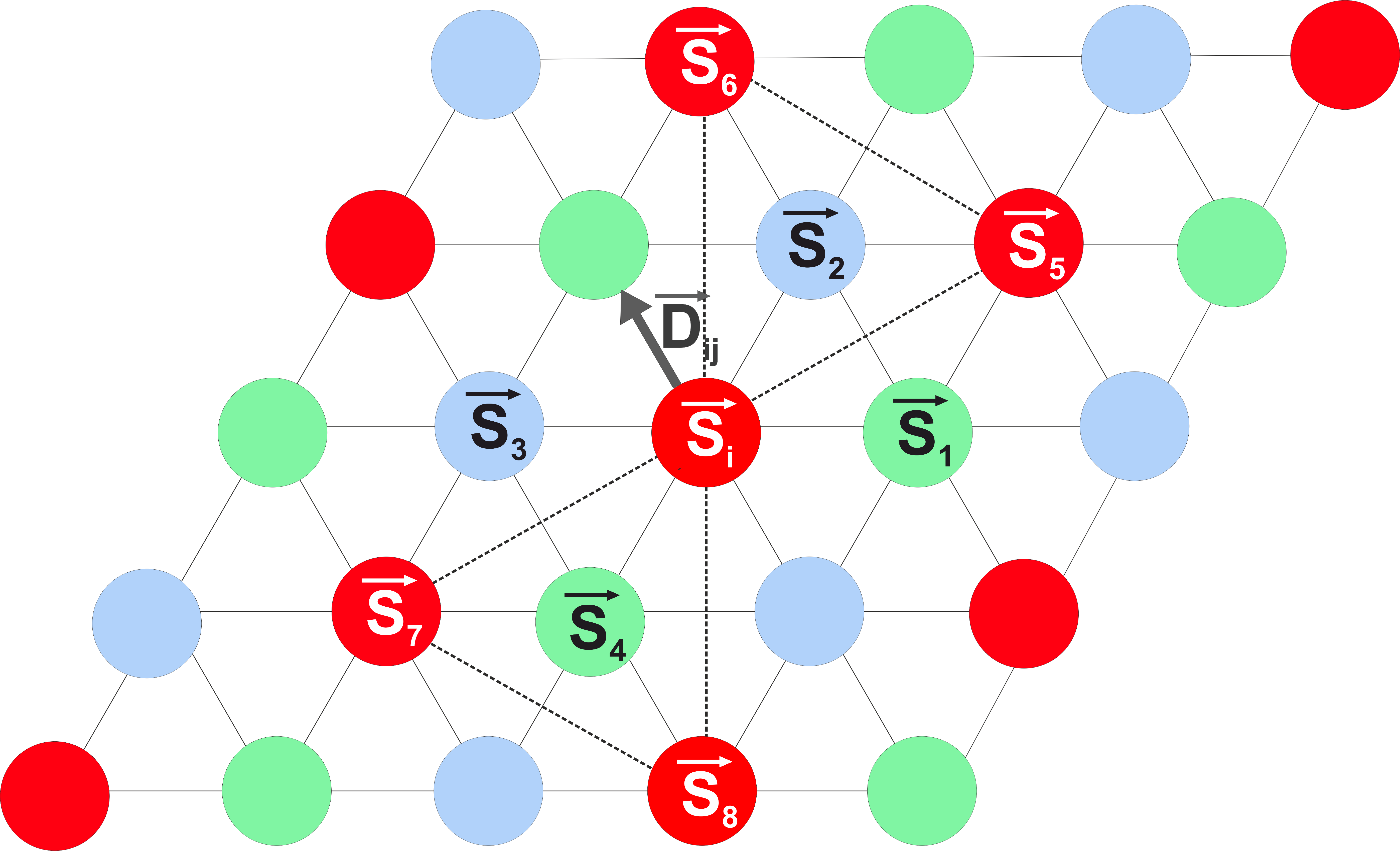}}
\caption{Three-sublattice decomposition of the triangular lattice. $\vec{S_i}$ is the central spin, $\vec{S_1},\hdots,\vec{S_4}$ and $\vec{S_5},\hdots,\vec{S_8}$ are the spins involved in calculation of the local chirality and the skyrmion number, respectively, and $\vec{D}_{ij}$ represents the Dzyaloshinskii-Moriya vector.}
\label{fig:latt}
\end{figure}

We consider the classical Heisenberg antiferromagnet on a triangular lattice with the Hamiltonian
\begin{equation}
\mathcal{H} = - J \sum_{\langle i,j \rangle}\vec{S_{i}}\cdot
\vec{S_{j}} + \sum_{\langle i,j \rangle}\vec{D_{ij}}
\cdot\Big [\vec{S_{i}}\times\vec{S_{j}} \Big] - h\sum_i S_i^z + A\sum_i (S_i^z)^2,
\label{hamilt}
\end{equation}

where $\vec{S_i}$ is a classical Heisenberg spin (vector of unit length) at the $i$th site, $J < 0$ is the antiferromagnetic exchange coupling constant, $h$ is the external magnetic field applied perpendicular to the lattice plane (along the $z$ direction), $A$ is the single-ion anisotropy strength and $\langle i,j\rangle$ denotes the sum over nearest-neighbour spins. $\vec{D}_{ij}$ is the Dzyaloshinskii-Moriya vector with the orientation defined by the crystal symmetries. Following the work of Rosales~\emph{et~al.}~\cite{rosales2015three}, we chose it to point along the radius-vector $\vec{r}_{ij}=\vec{r}_i - \vec{r}_j$ connecting two neighbouring sites, i.e., $\vec{D}_{ij} = D\frac{\vec{r}_{i,j}}{|\vec{r}_{i,j}|}$ (Fig.~\ref{fig:latt}), thus leading to the formation of the Bloch-type skyrmions. The strength of the DMI is defined by the magnitude of the parameter $D$.

In order to construct the phase diagrams we implement the parallel tempering (PT) or replica exchange MC~\cite{swendsen1986replica}, which has proved to be a very helpful and reliable tool in studying systems with complex energy surfaces. The lattice consists of $N = L^2$ sites, where $L = 48 - 72$, and has periodic boundary conditions. We use $180-300$ temperatures (replicas), depending on the parameter values combinations and the system size, which are selected as manually-tuned geometric progression in order to keep adequate exchange rates at low temperatures and close to the phase boundaries and to have reasonable resolution at higher ones. Considering the massive computational requirements of the problem, we have implemented an algorithm parallelized on General Purpose Graphical Processing Units (GPGPU) using CUDA, which allowed to simulate all the replicas at different parameter values simultaneously. For each replica we use $7-10 \times 10^6$ MC sweeps (MCS) for equilibration and the same amount for calculating mean values. The replica swapping occurs after each Metropolis sweep through the whole lattice. The lifetime calculations are based on the standard Metropolis method. Each state of the system is recorded after the whole lattice is swept and, thus, the lifetime is measured in MCS. In these calculations we use up to $1.5 \times 10^7$ MCS.

In order to identify different phases and to determine the phase boundaries we calculate several basic quantities, such as the magnetization $m$, the magnetic susceptibility $\chi_m$, and the specific heat $c$, as follows:

\begin{equation}
m = \frac{\langle M \rangle}{N} = \frac{1}{N} \Big\langle \sum_i S_i^z\Big\rangle,
\label{magn}
\end{equation}

\begin{equation}
\chi_m = \frac{\langle M^2\rangle  - \langle M \rangle^2 }{NT},
\label{magnsus}
\end{equation}

\begin{equation}
c = \frac{\langle \mathcal{H}^2\rangle  - \langle \mathcal{H} \rangle^2 }{NT^2}, 
\label{heatcap}
\end{equation}
where $\langle \cdots \rangle$ denotes the thermal average.

For capturing the skyrmion state in addition to the above standard thermodynamic quantities we also compute the skyrmion chirality and the skyrmion number, the discretizations of a continuum topological charge~\cite{berg1981definition}, which reflect the number and the nature of topological objects present in the system. The topological charge of a single skyrmion is $\pm 1$ for the core magnetization $\pm |\vec{S}|$~\cite{zhang2020skyrmion}. The skyrmion chirality $\kappa$, the skyrmion number $q$, and the corresponding susceptibilities $\chi_{\kappa}$ and $\chi_{q}$ are defined as follows:

\begin{equation}
\kappa = \frac{\langle K \rangle}{N} = \frac{1}{8\pi N} \Big\langle \sum_i \Big( \kappa^{12}_{i} + \kappa^{34}_{i} \Big)\Big\rangle,
\label{chiral}
\end{equation}

\begin{equation}
q = \frac{\langle Q \rangle}{N_s} = \frac{1}{4\pi N_s} \Big\langle\Big|\sum_i \Big( A^{56}_{i} \sign(\kappa^{56}_{i}) +A^{78}_{i} \sign( \kappa^{78}_{i}) \Big)\Big|\Big\rangle,
\label{sknum}
\end{equation}

\begin{equation}
\chi_{\kappa}= \frac{\langle K^2\rangle  - \langle K \rangle^2 }{NT},  
\label{xisus}
\end{equation}

\begin{equation}
\chi_q = \frac{\langle Q^2\rangle  - \langle Q \rangle^2 }{N_sT},
\label{xiQsus}
\end{equation}
where $\kappa^{ab}_{i} = \vec{S_i}\cdot[\vec{S_a}\times\vec{S_b}]$ is a chirality of a triangular plaquette of three neighbouring spins and $A^{ab}_{i} = ||(\vec{S_a} - \vec{S_i})\times(\vec{S_b} - \vec{S_i})||/2$ is the area of the triangle spanned by those spins. The chirality is calculated for the whole lattice and the summation runs through all the spins with $\{\vec{S_a}, \vec{S_b}\}$ corresponding to  $\{\vec{S_1}, \vec{S_2}\}$ and $\{\vec{S_3}, \vec{S_4}\}$ in Fig. \ref{fig:latt}, whereas the skyrmion number is calculated for each of the three sublattices, hence $N_s$ in Eqs.~(\ref{sknum}) and (\ref{xiQsus}) is the number of sites in each of the sublattices, $N_s = L^2/3$, and the triangular plaquette for the local quantities is formed by the neighbouring spins of the given sublattice $\{\vec{S_i}, \vec{S_5}, \vec{S_6}\}$ and $\{\vec{S_i}, \vec{S_7},\vec{S_8}\}$ in Fig.~\ref{fig:latt}. Spins in both cases are taken in counter-clockwise fashion to keep the sign in accordance with the rules in ~\cite{berg1981definition}.
 
To calculate the lifetime of the skyrmions, following the work of Rosales~\emph{et~al.}~\cite{rosales2015three}, we start with the spin configuration corresponding to the skyrmion lattice state at certain non-zero field value $h$ and then we quench the field to some final value $h_f << h$. We let the system relax and record the evolution of the chirality until no traces of the skyrmion phase remain in the system. In principle, such a situation could be detected as by the vanishing chirality. However, due to the finiteness of the system the chirality remains small but finite even after the skyrmions have collapsed. Therefore, we arbitrarily applied the cut-off corresponding to the (last) inflexion point of the chirality evolution curve. 

\section{Results}
\subsection{Phase diagrams}
It has been shown that the model without the single-ion anisotropy features four ordered phases: the helical (HL) phase, the coplanar up-up-down (UUD), the canted V-like (VL), and the skyrmion lattice (SkX) phases~\cite{rosales2015three,mohylna2021formation}. As the DMI increases the SkX phase appears at very low temperatures and intermediate fields, wedged between the HL and VL phases, and its area grows and extends to higher temperatures. The HL-SkX phase boundary at sufficiently low temperatures displays discontinuous first-order character, which changes to the second-order behaviour at the tricritical point.

\subsubsection{Thermodynamic quantities}
\begin{figure}[t!]
\centering
\subfigure{\includegraphics[scale=0.28,clip]{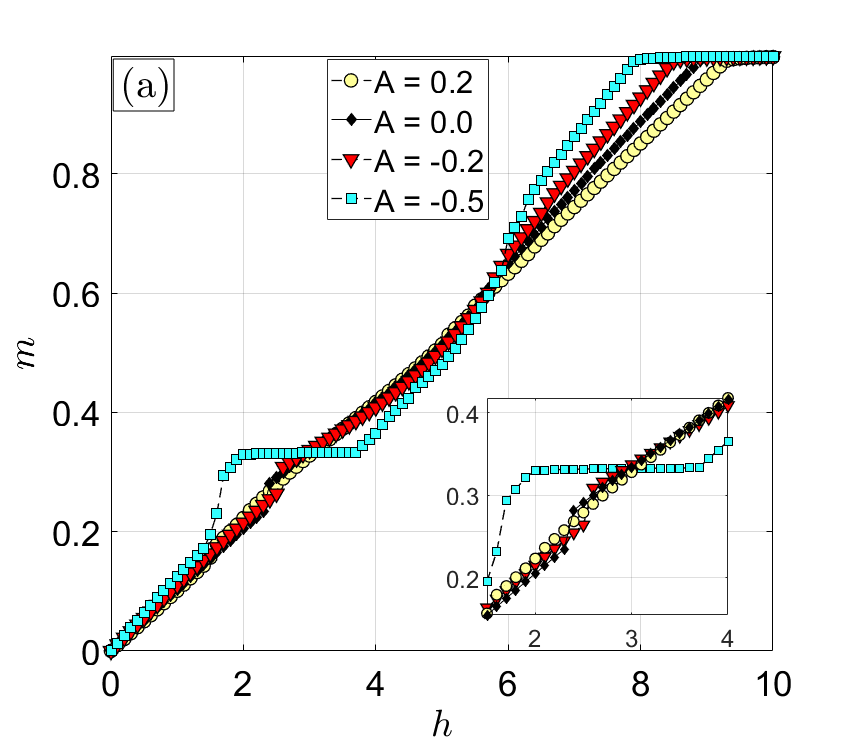}\label{fig:mz_D050}}
\subfigure{\includegraphics[scale=0.28,clip]{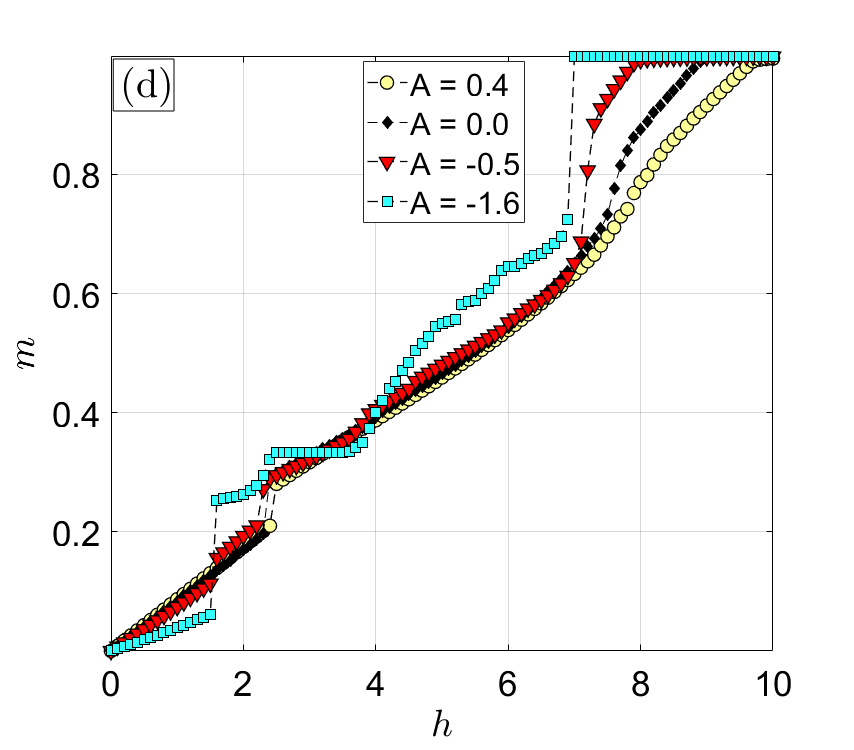}\label{fig:mz_D100}}\vspace{-3mm}\\
\subfigure{\includegraphics[scale=0.28,clip]{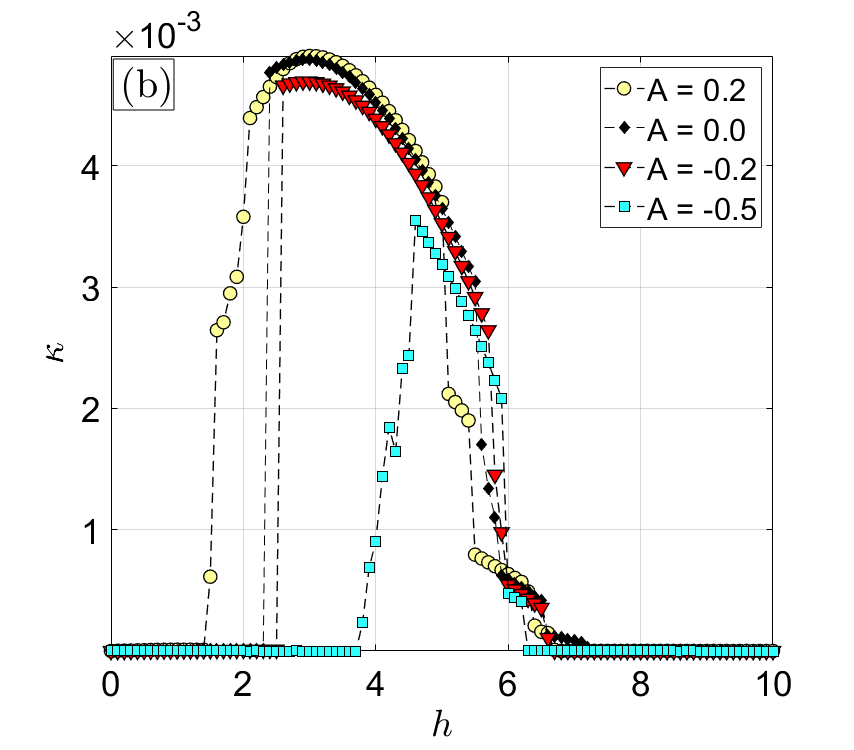}\label{fig:xiL_D050}}
\subfigure{\includegraphics[scale=0.28,clip]{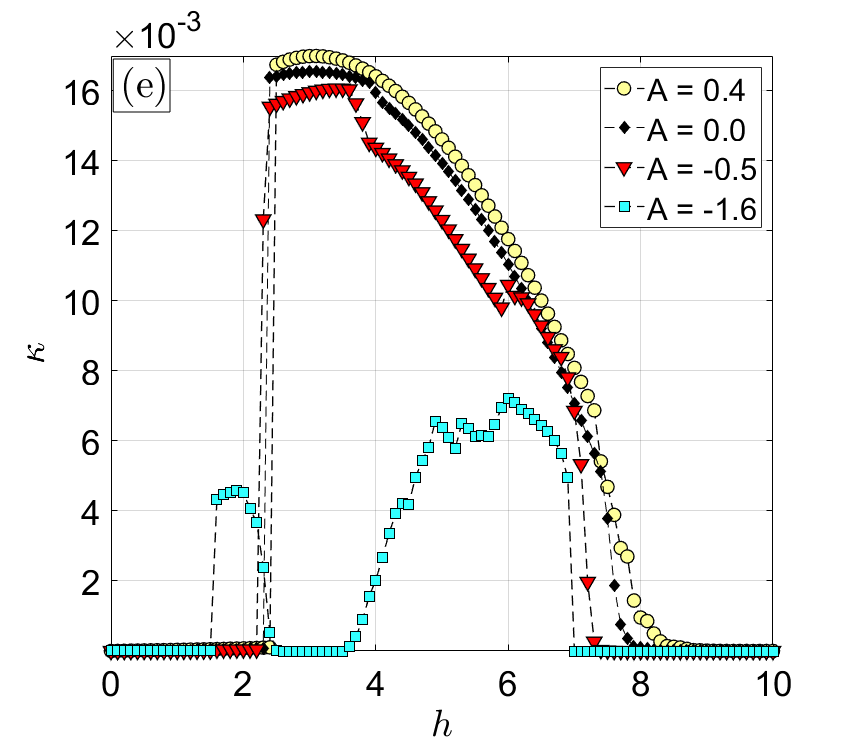}\label{fig:xiL_D100}}\vspace{-3mm}\\
\subfigure{\includegraphics[scale=0.28,clip]{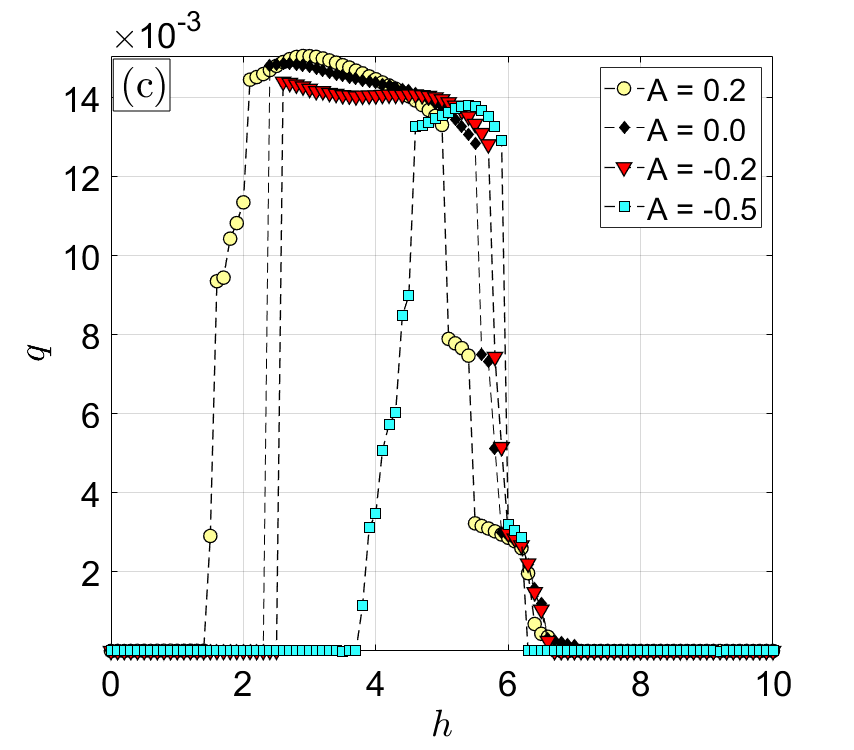}\label{fig:xiQ_D050}}
\subfigure{\includegraphics[scale=0.28,clip]{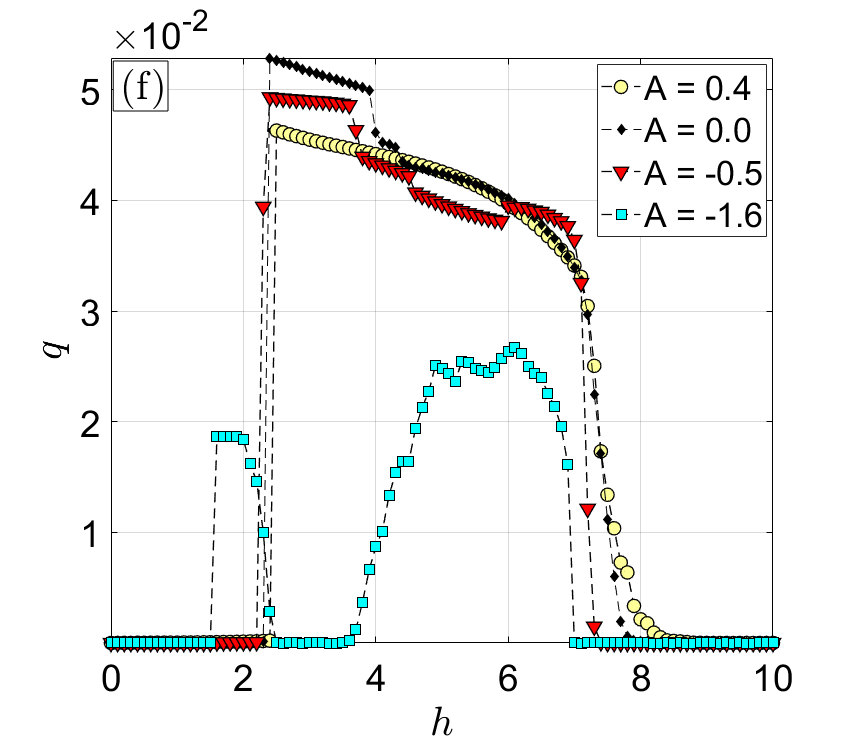}\label{fig:xiQ_D100}}
\caption{Field dependencies of (a, d) the magnetization, (b, e) the chirality and (c, f) the skyrmion number, for D = 0.5 (left panels) and D = 1.0 (right panels) with $T = 0.01$ and $L = 48$.}
\label{fig:quant}
\end{figure}

To demonstrate effects of the single-ion anisotropy, first we present dependencies of the above introduced thermodynamic quantities, for some representative parameter values. In particular, in Fig.~\ref{fig:quant} field dependencies of the order parameters $m,\kappa$ and $q$ are plotted for different anisotropy values with the fixed $T = 0.01$ and the DMI parameter corresponding to $D = 0.5$ (left panels) and $D = 1$ (right panels). To better illustrate the effect of the single-ion anisotropy on the behaviour of these quantities, we also include the isotropic case of $A=0$ (black symbols).

For the isotropic case the magnetization has been shown to display a monotonically increasing dependence with a jump at the phase transition between the HL and SkX phases at $h \approx 2.4$ and less conspicuous anomaly at the SkX-VL phase transition above $h \approx 6$, before leveling off at $h \approx 9$ corresponding to the onset of the fully polarized paramagnetic (P) phase~\cite{rosales2015three,mohylna2021formation}. The easy-plane anisotropy overall tends to smoothen the magnetization curves and suppress the discontinuous behaviour. On the other hand, the easy-axis anisotropy makes the magnetization process more Ising-like~\cite{metcalf1973phase} with the discontinuities more pronounced and a plateau corresponding to one third of the saturation value showing within some field range, thus indicating the appearance of the UUD phase. For small $D$ the curves resemble that for the anisotropic Heisenberg model without the DMI~\cite{yun2015classical}, however, larger DMI results in a more complex behaviour with possible multiple steps, as shown in the top row of Fig.~\ref{fig:quant} for $D=1$.


More suitable quantities for determining the SkX phase boundaries are the chirality (the middle row of Fig.~\ref{fig:quant}) and the skyrmion number (the bottom row of Fig.~\ref{fig:quant}), being the order parameters of that phase. For smaller DMI and easy-axis anisotropy, such as $D = 0.5$ and $A=-0.2$, one can notice that the magnetization jump at $h \approx 2$ coincides with a sharp increase of both the chirality and the skyrmion number, meaning the first-order phase transition to the skyrmion state. However, with the anisotropy increased to $A = -0.5$ the magnetization jumps at the transition to the UUD phase with zero chirality and zero skyrmion number, which extends up to $h \approx 4$, thus squeezing the range of the SkX phase. On the other hand, a small easy-plane anisotropy, such as $A=0.2$, tends to slightly extend the field range of the SkX phase by shifting the lower bound to lower fields. 

\begin{figure}[t!]
\centering
\subfigure{\includegraphics[scale=0.28,clip]{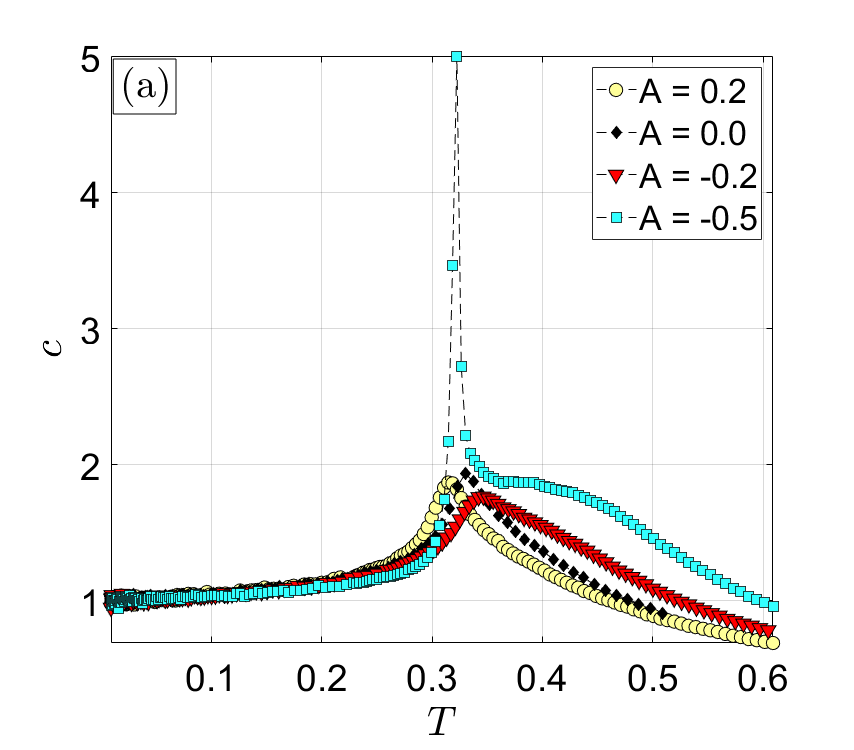}\label{fig:cv_D050}}
\subfigure{\includegraphics[scale=0.28,clip]{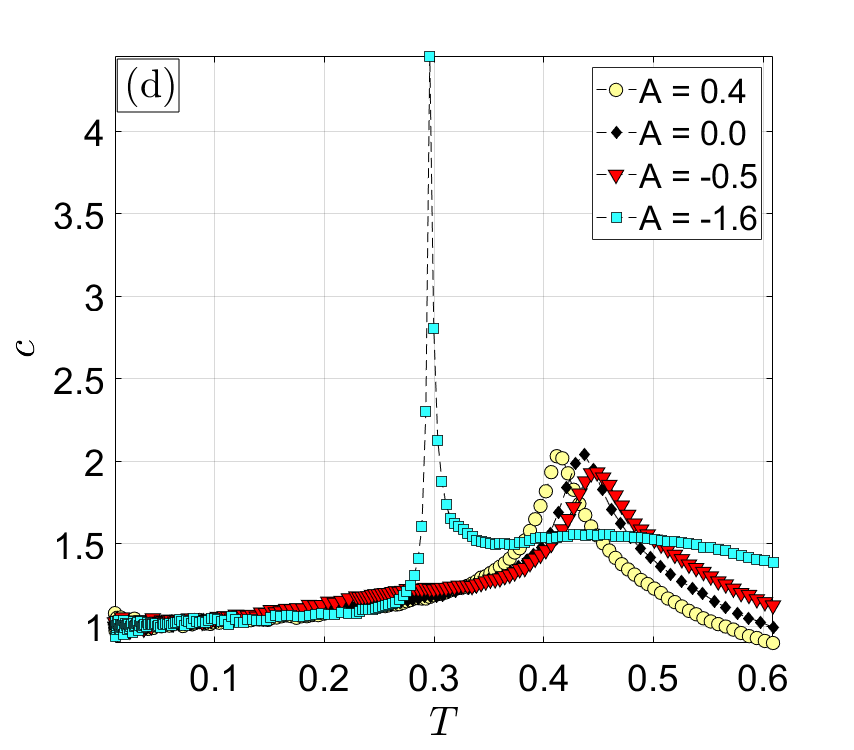}\label{fig:cv_D100}}\vspace{-3mm}\\
\subfigure{\includegraphics[scale=0.28,clip]{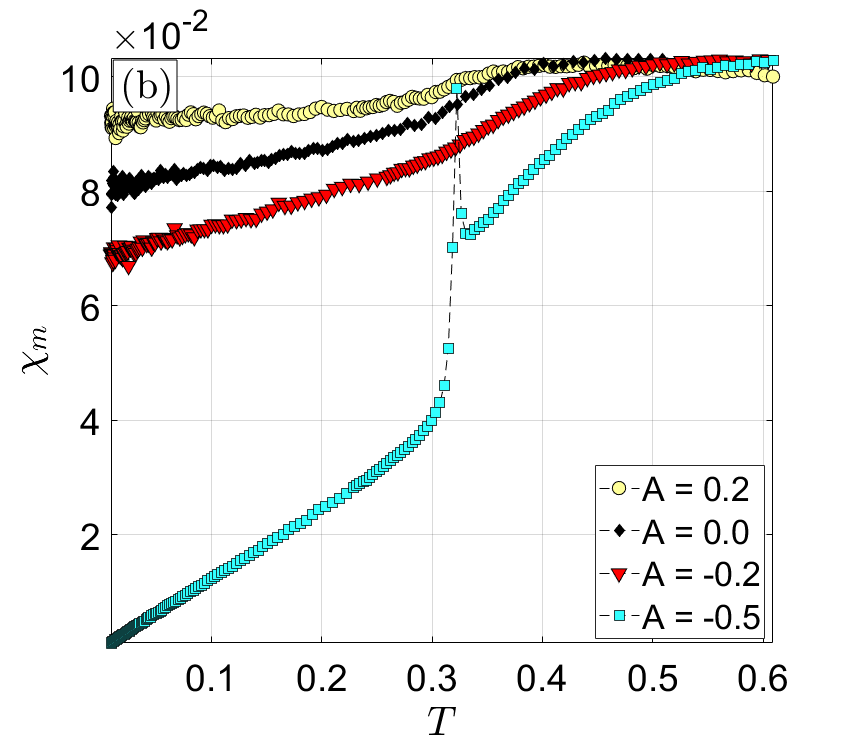}\label{fig:mzSus_D050}}
\subfigure{\includegraphics[scale=0.28,clip]{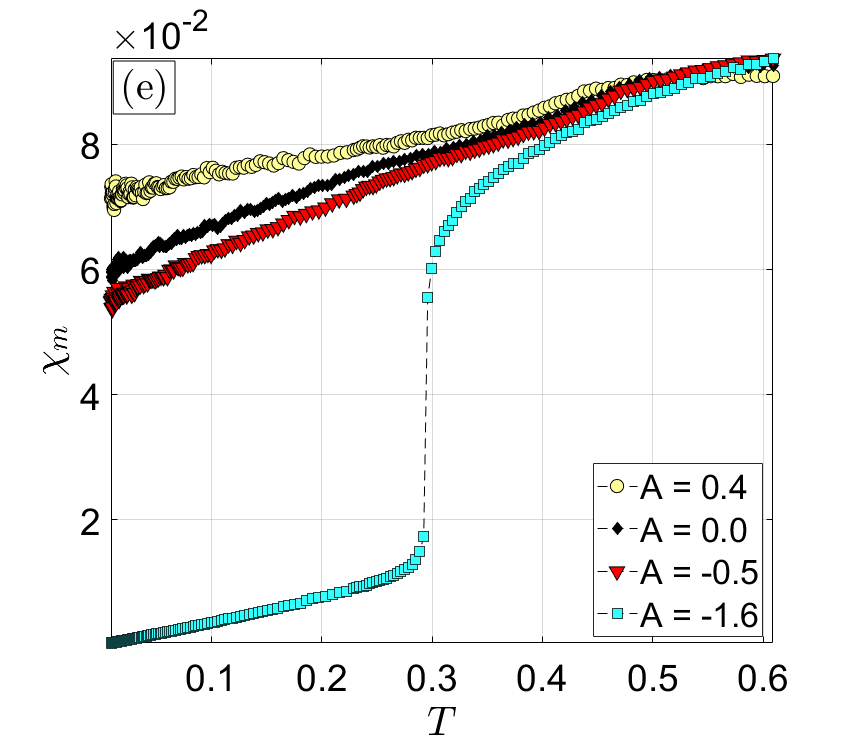}\label{fig:mzSus_D100}}\vspace{-3mm}\\
\subfigure{\includegraphics[scale=0.28,clip]{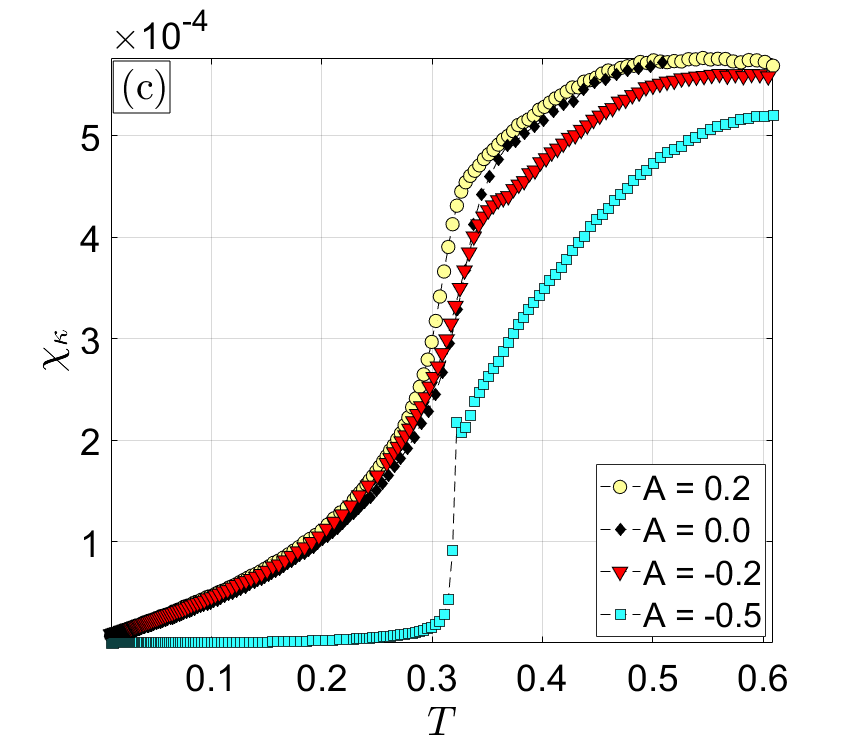}\label{fig:xiQSus_D050}}
\subfigure{\includegraphics[scale=0.28,clip]{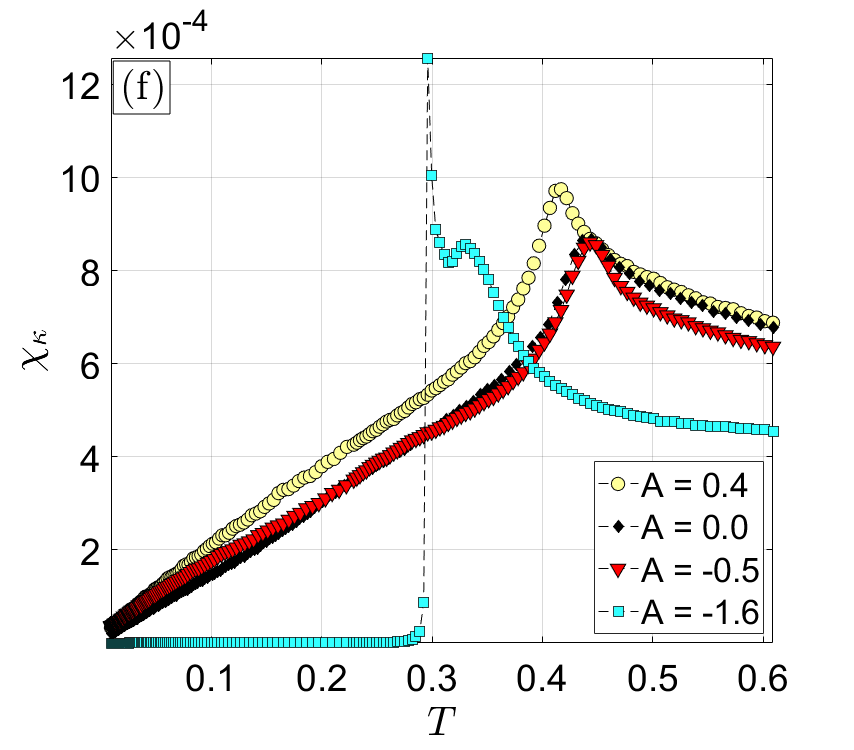}\label{fig:xiQSus_D100}}
\caption{Temperature dependencies of (a, d) the specific heat, (b, e) the magnetic susceptibility and (c, f) the chiral susceptibility, for $h = 3.0$ and $D = 0.5$ (left panels) and $D = 1.0$ (right panels).}
\label{fig:resp}
\end{figure}

For bigger DMI value, such as $D = 1$, with the increasing easy-axis anisotropy within the SkX phase one can observe depressions formed in both the chirality and the skyrmion number dependencies (see the curves for $A = -0.5$), related to the change of the skyrmions' size and consequently their number. With further increase of the easy-axis anisotropy, an interesting reentrant behaviour occurs for the SkX phase. In particular, as demonstrated by the behaviour of the chirality and the skyrmion number for $A = -1.6$, with the increasing field the system first enters the SkX phase and stays in it within some field range around $h = 2$, then it crosses to the UUD phase and again returns to the SkX phase at $h \approx 4$, before reaching the P phase above $h \approx 7$. It is worth to notice that the low-field part of the SkX phase for the easy-axis anisotropy occurs at the fields lower than the compact SkX phase for the zero or easy-plane anisotropy. Thus, in terms of the minimum field intensity necessary for the emergence of the SkX phase, the effect of the increasing DMI in the systems with easy-axis and easy-plane anisotropies is opposite.


Some of the phase boundaries are more conveniently established using various response functions, as shown in Fig.~\ref{fig:resp}. For example, for moderate values of $A$ the specific heat (upper row) displays distinct peaks at the SkX-P phase transition. However, when the easy-axis anisotropy exceeds some threshold value (e.g., $A=-0.5$ for $D=0.5$ or $A=-1.6$ for $D=1$) the phase transition from the newly created UUD phase to the SkX phase is accompanied with the spike-like peak, signifying the first-order transition, while the second peak at higher temperatures related to the SkX-P transition becomes rounded and less defined. 

\subsubsection{Phase diagram evolution}
\begin{figure}[t!]
\centering
\subfigure{\includegraphics[scale=0.28,clip]{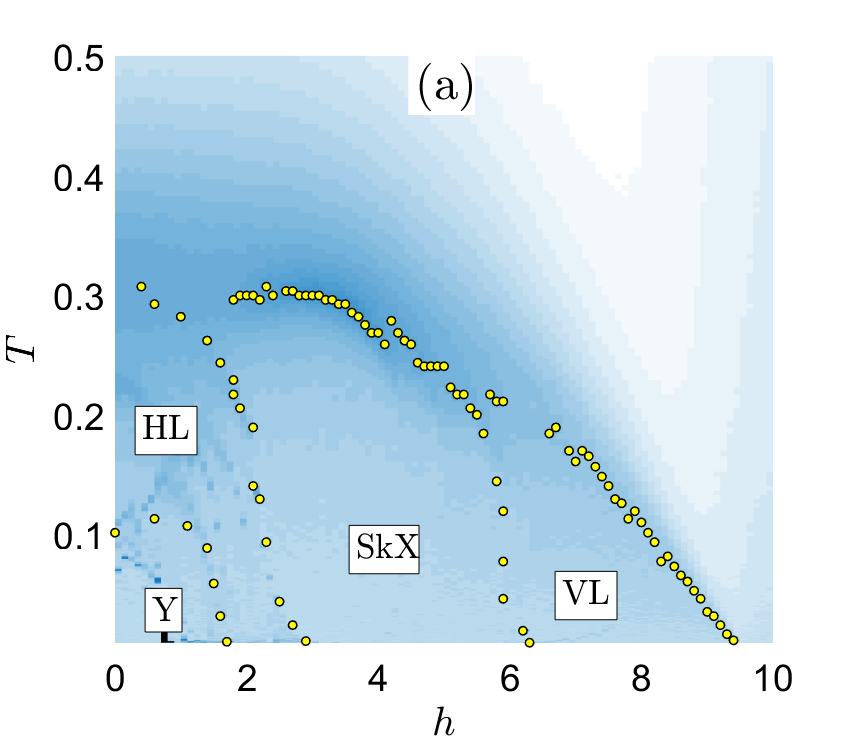}\label{fig:D050A0.3}}
\subfigure{\includegraphics[scale=0.28,clip]{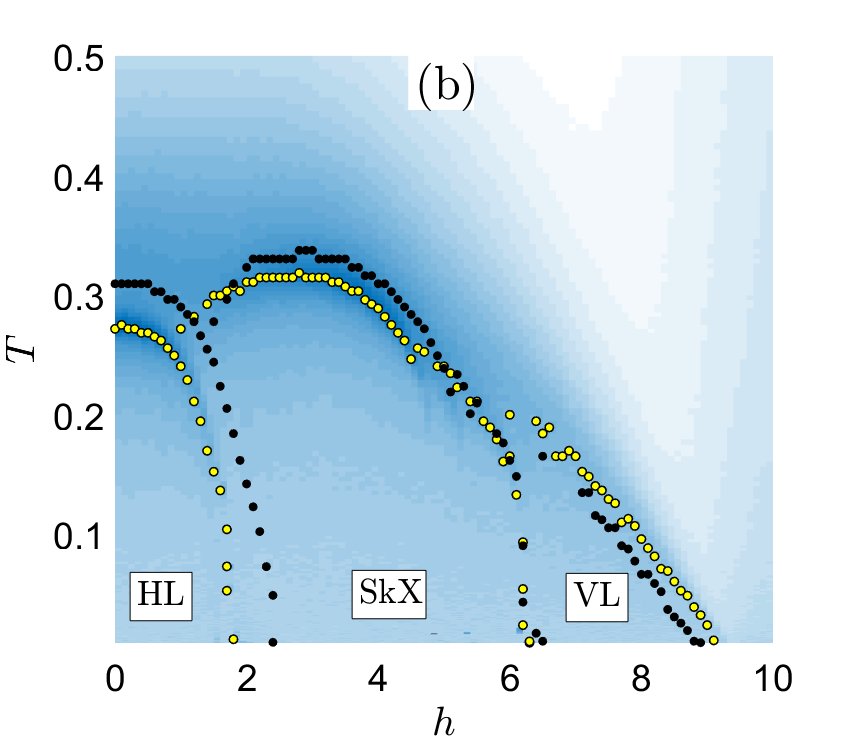}\label{fig:D050A0.2}}\vspace{-3mm}\\
\subfigure{\includegraphics[scale=0.28,clip]{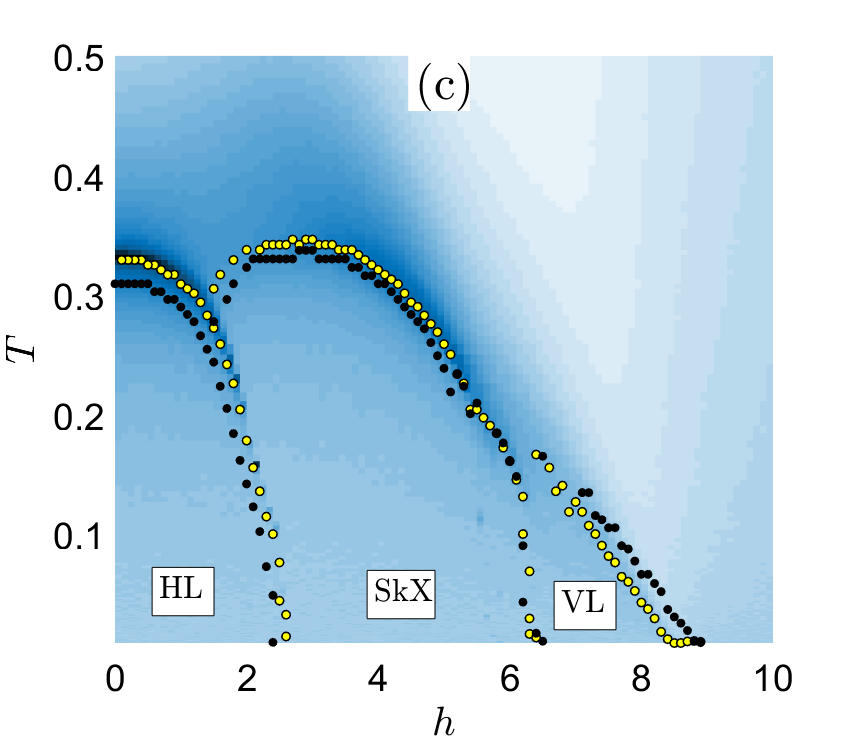}\label{fig:D050A-0.2}}
\subfigure{\includegraphics[scale=0.28,clip]{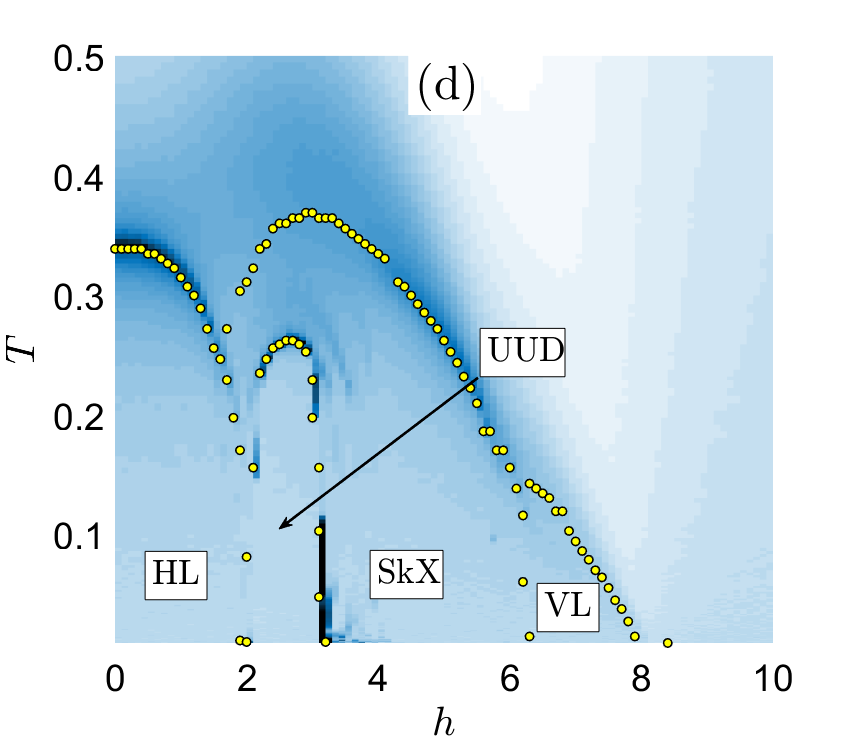}\label{fig:D050A-0.4}}\vspace{-3mm}\\
\subfigure{\includegraphics[scale=0.28,clip]{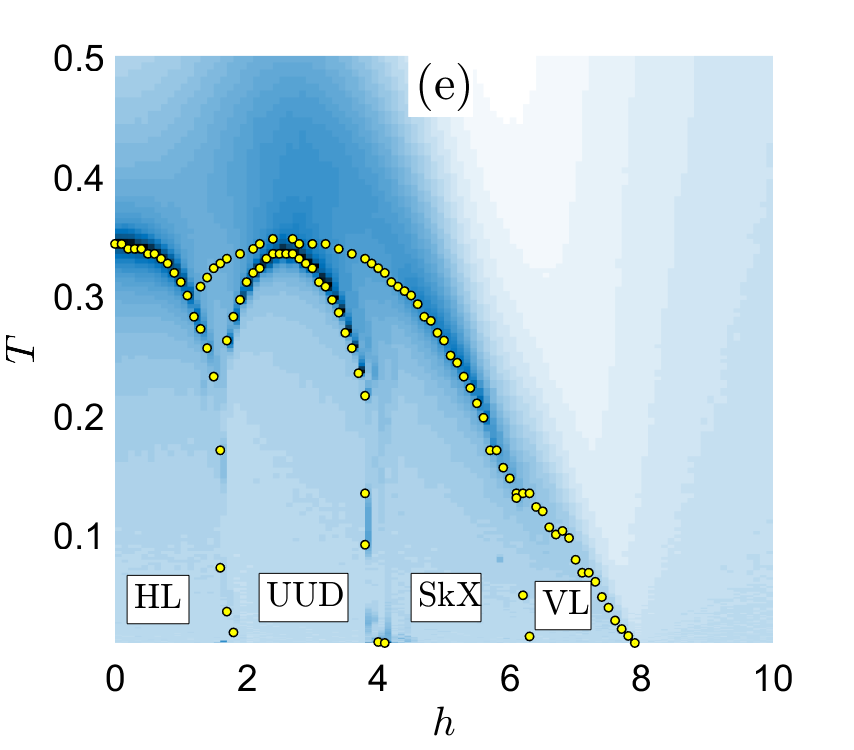}\label{fig:D050A-0.5}}
\subfigure{\includegraphics[scale=0.28,clip]{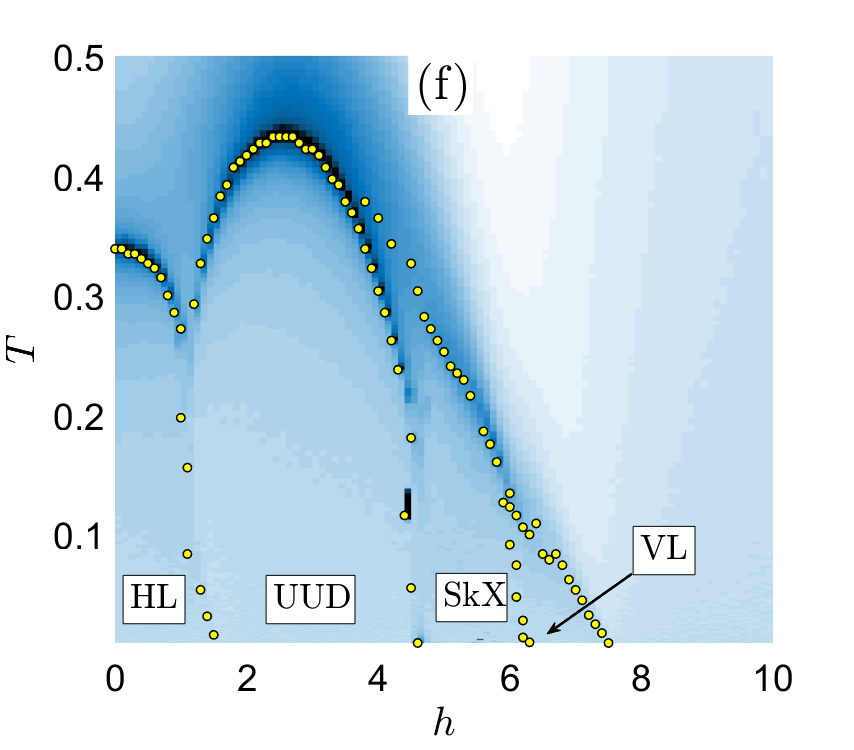}\label{fig:D050A-0.7}}
\caption{Phase diagrams in $T-h$ plane for $D = 0.5$ with (a) $A = 0.3$, (b) $A = 0.2$, (c) $A = -0.2$, (d) $A = -0.4$, (e) $A = -0.5$ and (f) $A = -0.7$. The black circles in (b) and (c) panels represent the isotropic $A = 0$ case. The background colour corresponds to the specific heat magnitude.}
\label{fig:PB_D050}
\end{figure}

\begin{figure}[t!]
\centering
\subfigure{\includegraphics[scale=0.28,clip]{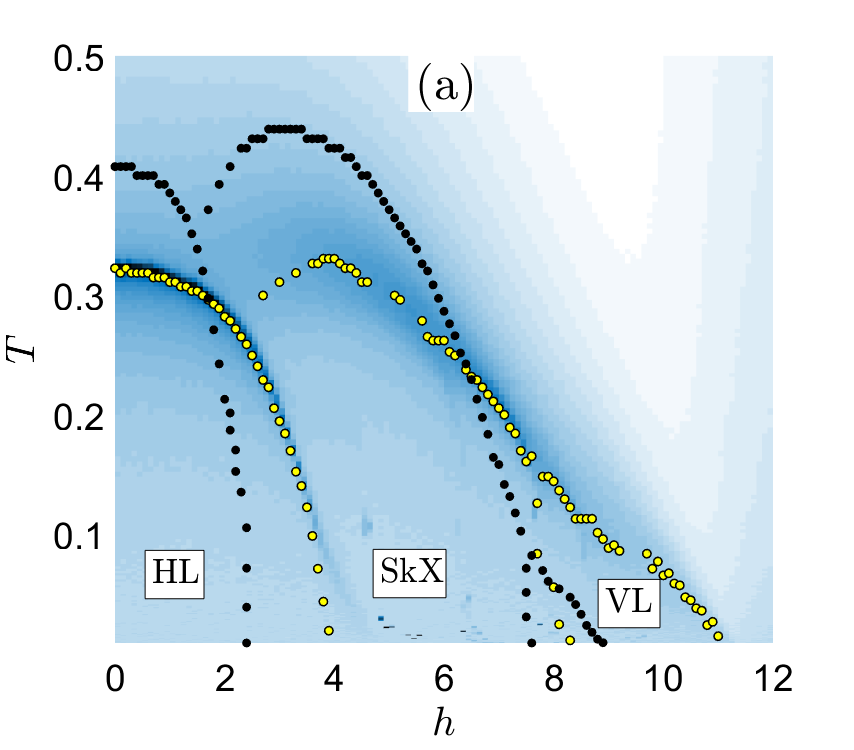}\label{fig:D100A1.2}}
\subfigure{\includegraphics[scale=0.28,clip]{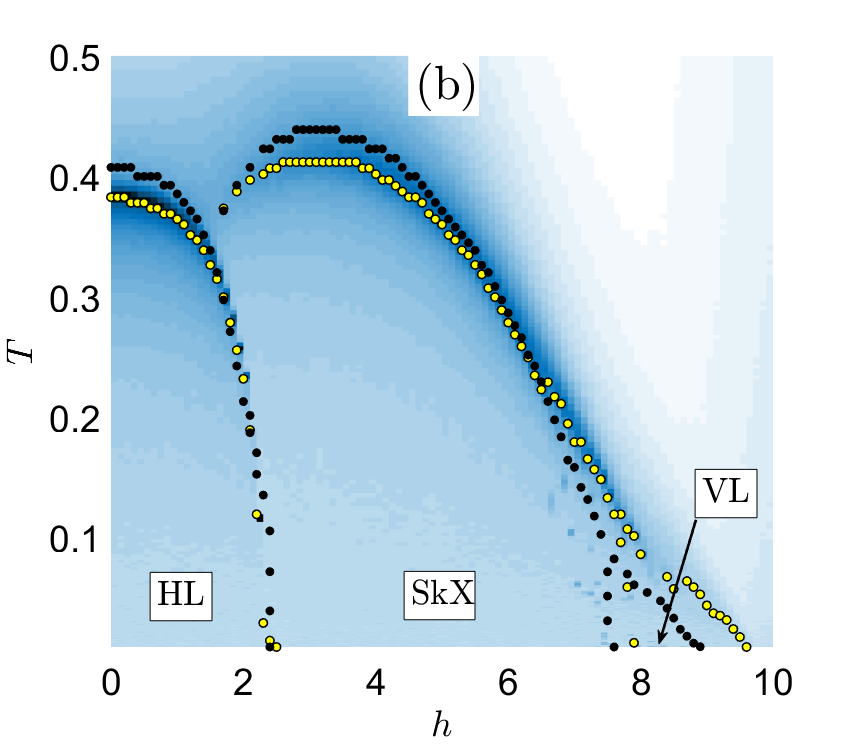}\label{fig:D100A0.4}}\vspace{-3mm}\\
\subfigure{\includegraphics[scale=0.28,clip]{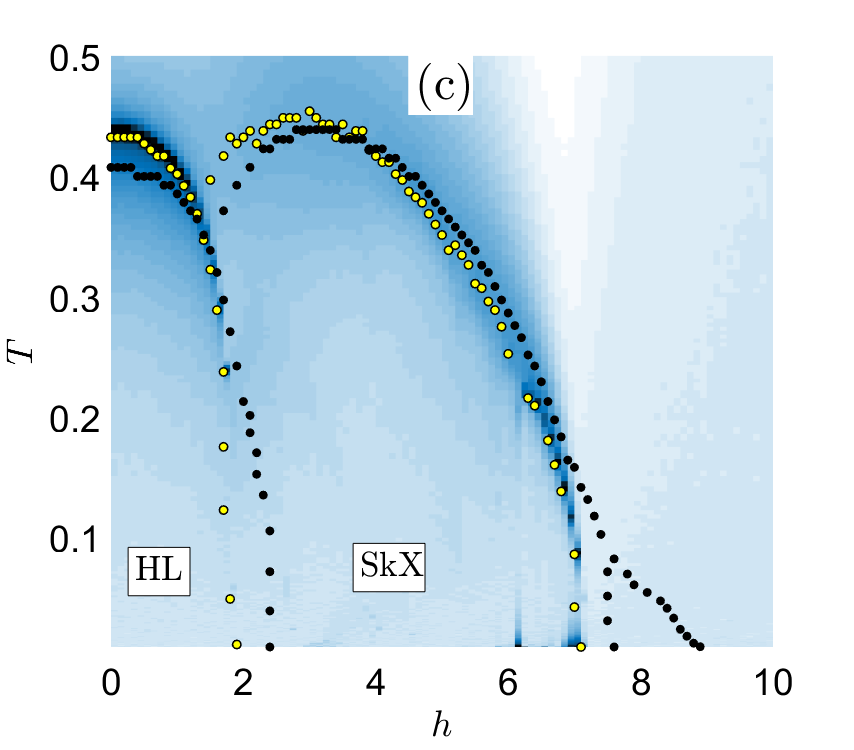}\label{fig:D100A-1.0}}
\subfigure{\includegraphics[scale=0.28,clip]{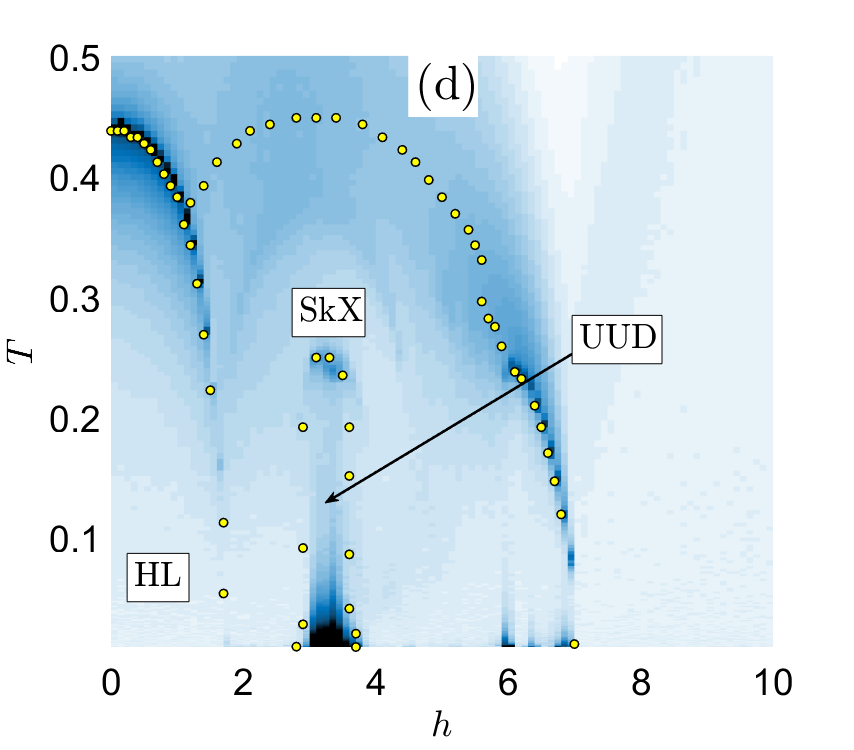}\label{fig:D100A-1.5}}\vspace{-3mm}\\
\subfigure{\includegraphics[scale=0.28,clip]{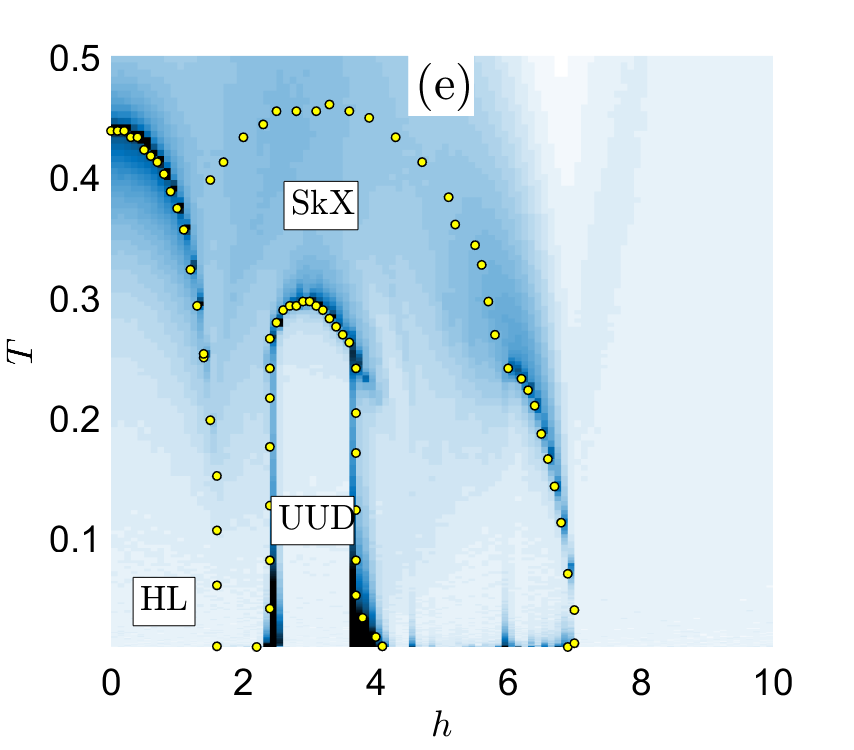} \label{fig:D100A-1.6}}
\subfigure{\includegraphics[scale=0.28,clip]{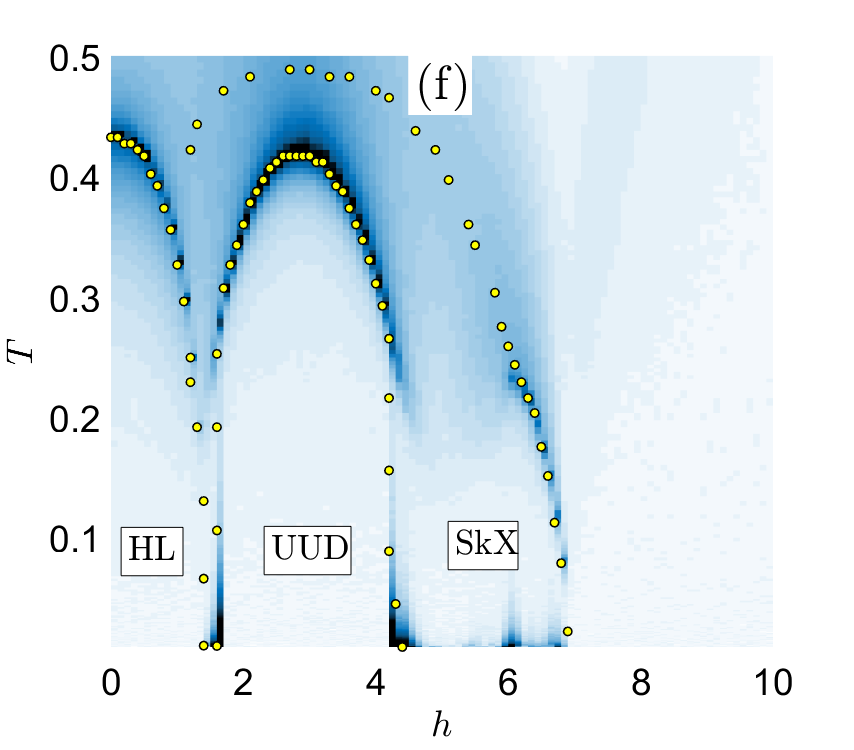}\label{fig:D100A-2.0}}
\caption{Phase diagrams in $T-h$ plane for $D = 1.0$ with (a) $A = 1.2$, (b) $A = 0.4$, (c) $A = -1.0$, (d) $A = -1.5$, (e) $A = -1.6$ and (f) $A = -2.0$. The black circles in (b) and (c) panels represent the isotropic $A = 0$ case. The background colour corresponds to the specific heat magnitude.}
\label{fig:PB_D100}
\end{figure}

Below we present the phase diagrams in $T-h$ plane for two selected values of $D=0.5$ (Fig.~\ref{fig:PB_D050}) and $D=1$ (Fig.~\ref{fig:PB_D100}) and various values of $A$. They were constructed using locations of the peaks in the evaluated response functions, $c,\chi_{m}$, and $\chi_q$, the magnetization $m$ and the order parameters of the SkX phase, $\kappa$ and $q$, as well as by visual inspection of spin snapshots. The calculations were carried out for a fixed lattice size of $L = 48$ and, therefore, strictly speaking, the obtained results are pseudo phase diagrams showing only approximate locations of the phase boundaries. The estimated (pseudo) transition temperatures, shown as yellow circles, are overlaid on the specific heat background with dark blue colour corresponding to higher (peak) values. To better visualize effects of the single-ion anisotropy, for the cases with relatively small (both easy-axis and easy-plane) anisotropy we also added the phase diagrams corresponding to $A=0$, shown in black.

The topology of the phase diagram of the isotropic ($A = 0$) case has been already discussed in several works~\cite{mohylna2021formation, rosales2015three, osorio2017composite} and also briefly introduced in the first paragraph of this section. The phase diagrams for the isotropic case with $D=0.5$ and $D=1$ are included in Figs.~\ref{fig:PB_D050} and~\ref{fig:PB_D100}, respectively. For $D=0.5$ the phase formed at lower fields up to $h \approx 2.4$ is the HL phase with spin stripes rotating in $x-y$ plane. The sharp jump of the skyrmion number from zero to finite values at $h \approx 2.4$ (included in Fig.\ref{fig:quant}) signifies the first-order phase transition to the AFM SkX phase with three skyrmion sublattices, which persists up to the field $h \approx 6$. Further increase of the field brings the system to the VL phase, which is a variation of the canted 2:1 (V) phase in the Heisenberg triangular AFM without the DMI~\cite{seabra2011phase}, obeying the condition for the three neighbouring spins on the triangular plaquette: $\vec{S_1}+\vec{S_2}+\vec{S_3} = \vec{h}/3$~\cite{osorio2017composite,mohylna2021formation}.
 
For the case of $D=0.5$, one can observe that the inclusion of a relatively small easy-plane anisotropy ($A = 0.2$ in Fig.~\ref{fig:D050A0.2}) slightly decreases the temperature range of the SkX phase but, on the other hand, it extends its boundaries to lower fields at the expense of the HL phase. However, further increase of the easy-plane anisotropy (such as $A = 0.3$ in Fig.~\ref{fig:D050A0.3}) diminishes both the SkX and HL phases by pushing the field lower bound of the former to higher values and by replacing the part of the latter in the low-field region with the so-called Y ($120^{\circ}$) phase. On the other hand, a small easy-axis anisotropy marginally extends the temperature range of the SkX but at the same time it tends to shrink its field range (see Fig.~\ref{fig:D050A-0.2} for $A = -0.2$). Upon further increase of the easy-axis anisotropy the phase diagram topology changes. As shown in Fig.~\ref{fig:D050A-0.4} for $A = -0.4$, the UUD phase is formed in a narrow field region between the HL and SkX phases and with the increasing anisotropy it grows quite rapidly. For $A = -0.5$ it already occupies a major part of the plane previously taken by the SkX phase and at $A = -0.7$ only a small fraction of it remains at relatively high fields around $h = 5$. The growing UUD phase also squeezes the field range of the HL phase. This behaviour of the UUD phase with the increasing easy-axis anisotropy is consistent with that of the anisotropic Heisenberg model without the DMI~\cite{yun2015classical}, in which the UUD phase extends at the cost of the canted 2:1 phase from the high-field and, albeit much less markedly, the Y phase from the low-field sides. 

In the case of higher DMI, such as $D = 1.0$ in Fig.~\ref{fig:PB_D100}, the effect of the single-ion anisotropy appears somewhat different. A small easy-plane anisotropy (see Fig.~\ref{fig:D100A0.4} for $A=0.4$) reduces the temperature window of the SkX (and also HL) phase, as for the moderate DMI, but it only slightly shifts the upper limit of the field window at low temperatures, while the lower limit stays about the same. Further increase in $A$ (see Fig.~\ref{fig:D100A1.2} for $A=1.2$) results in additional lowering of the SkX-P and HL-P transition temperatures, shrinking the field window of the SkX phase and shifting it (particularly the lower bound) to higher values. The VL phase extends to higher temperatures and larger field values. On the other hand, the effects of the increasing easy-axis anisotropy are quite opposite. For moderate negative values of $A$ the SkX-P and HL-P transition temperatures slightly increase and the field window of the SkX phase shifts to smaller values but its width remains about the same. Both the temperature and field windows of the VL phase shrink. Below $A \approx -0.5$ the VL phase disappears and the fully polarized phase sets in for $h \gtrapprox 7$ (see Fig.~\ref{fig:D100A-1.0} for $A=-1$). At still larger easy-axis anisotropy ($A \approx -1.5$) the UUD phase appears inside the SkX phase and with the decreasing $A$ it broadens both the temperature and field ranges of its existence at the expense of the SkX phase.



\subsection{Skyrmion lifetime}

In the following we address the issue of the skyrmions stability in zero or small magnetic fields and the influence of different parameters on it. To do so we followed the approach adopted by Rosales~\emph{et~al.}~\cite{rosales2015three} and performed the Metropolis MC simulations initialized with a skyrmion configuration obtained at thermal equilibrium corresponding to different parameter values within the SkX phase and then quenched the initial magnetic field $h$ to some final value $h_f$, which is set either to zero or some other (smaller) value, corresponding to the HL phase. Then we let the system thermalize to the equilibrium. To monitor the the time evolution of the SkX states and their eventual collapse, we recorded time series of the SkX phase order parameter, i.e., the skyrmion chirality $\kappa$, as well as some other quantities, such as the magnetization $m$. The obtained data were configurationally averaged over 300 replicas obtained from independent MC runs initialized with the same configuration.

\begin{figure}[t!]
\centering
\subfigure{\includegraphics[scale=0.35,clip]{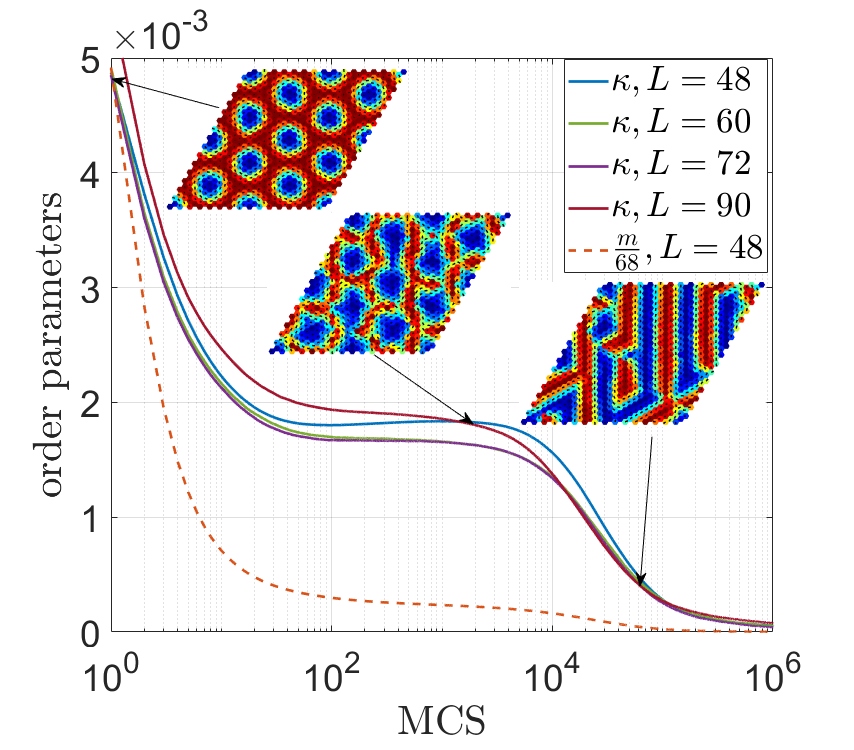}\label{fig:xi_mz_LT_D050h3}}
\caption{Solid curves show time evolution of the skyrmion chirality $\kappa$ from the initial state corresponding to the SkX phase after quenching the external field to zero, for different lattice sizes and the parameters $D = 0.5$, $T = 0.009$, and $h = 3.2$. The broken curve shows the corresponding time evolution of the magnetization $m$ for $L=48$. The snapshots illustrate the spin textures in different stages of the relaxation process.} 
\label{fig:LT_mz_xi}
\end{figure}

In Fig.~\ref{fig:LT_mz_xi} we present the typical behaviour of the skyrmion lifetime in terms of the number of the MCS needed to bring the skyrmion chirality to zero, shown for the parameter values $D = 0.5$, $T = 0.009$, and $h = 3.2$. We also considered different lattice sizes but no significant finite-size effects could be observed. Typically, the evolution curves show a rather steep decrease during the initial MCS before they level off at some finite value around which they persist over another tens of thousands of MCS. Within this impressively long time, corresponding to the plateau-like section of the skyrmion chirality evolution curve, the SkX phase still persists. As demonstrated by the representative one-sublattice spin snapshot taken after 2000 MCS, compared to the initial state the shape of skyrmions becomes a bit distorted, nevertheless, the skyrmion lattice features are is still very distinctive. After several tens of thousands of MCS the chirality drops to the values close to zero and, as shown in the last snapshot, spins unwind in helical-like stripes. In order to quantify the skyrmion lattice persistence, we arbitrarily chose the inflection point of the chirality evolution curve corresponding to its final decrease as the skyrmion lifetime, $\tau$. To compare it with the magnetic persistence we include in Fig.~\ref{fig:LT_mz_xi} (dashed curve) also the corresponding magnetization ($m/68$ in order to use the same scale) evolution curve. As one can see, the latter drops almost instantly to zero, thus even more underlining the remarkable skyrmion lattice stability. 


\begin{figure}[t!]
\centering
\subfigure{\includegraphics[scale=0.32,clip]{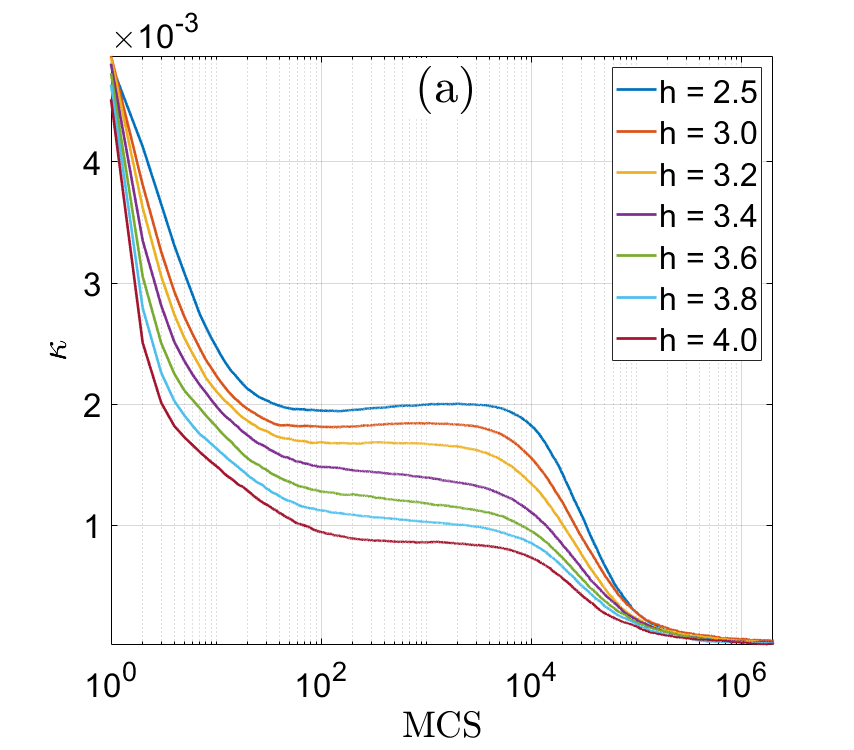} \label{fig:D_0.500T_0.009xiLTS}}
\subfigure{\includegraphics[scale=0.32,clip]{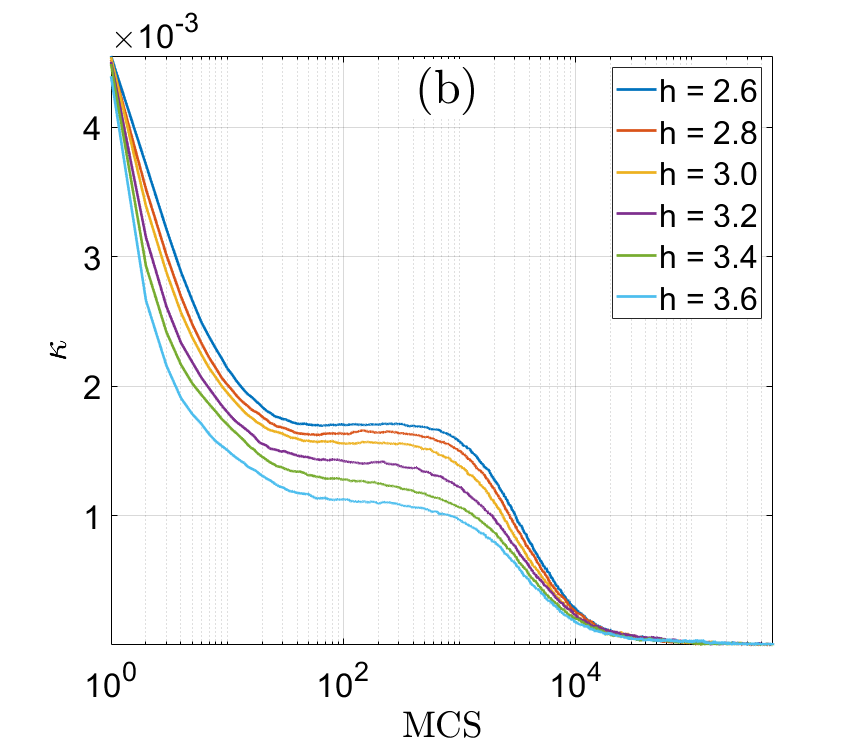}\label{fig:D_0.500T_0.050xiLTS}}\\
\subfigure{\includegraphics[scale=0.32,clip]{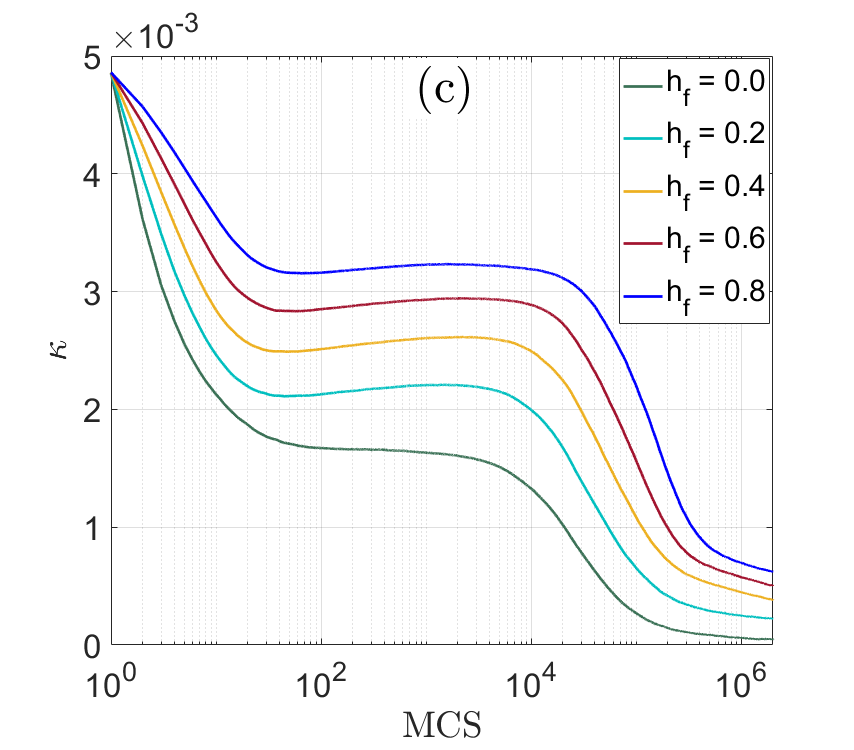} \label{fig:LT_H_pos}}
\subfigure{\includegraphics[scale=0.32,clip]{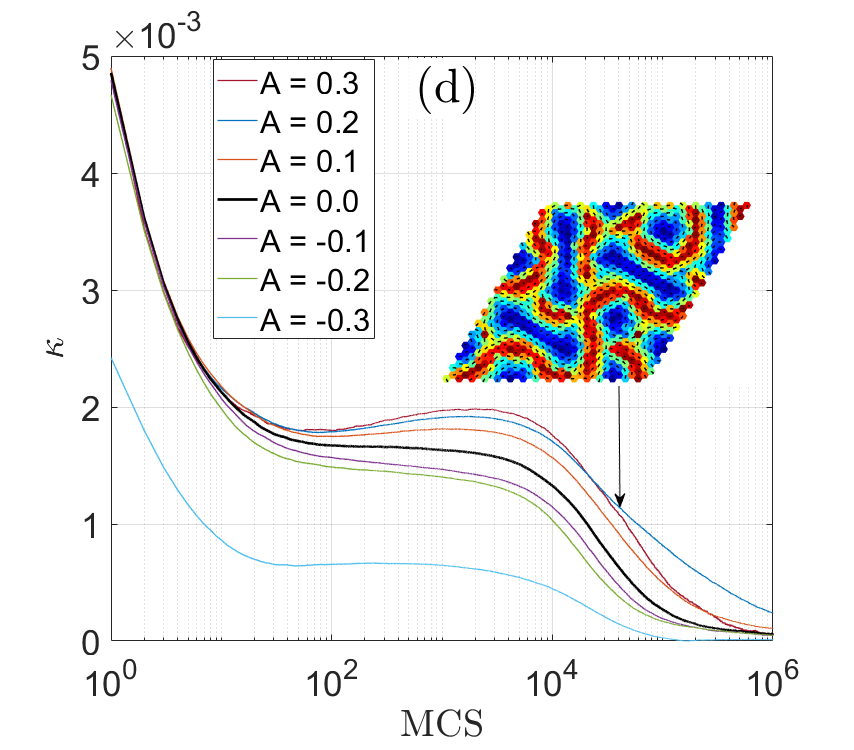}\label{fig:LT_with_ani_D050}}
\caption{Time evolution curves of the skyrmion chirality $\kappa$ from the initial state corresponding to the SkX phase after quenching the initial external field $h$ to the final value $h_f$, for (a) different $h$ with $A = 0$, $D = 0.5$, $h_f=0$ and $T = 0.009$, (b) different $h$ with $A = 0$, $D = 0.5$, $h_f = 0$ and $T = 0.05$, (c) different $h_f$ with $A=0$, $D = 0.5$, $h = 3.2$ and $T = 0.009$, and (d) different $A$ with $D = 0.5$, $h = 3.2$, $h_f = 0$ and $T = 0.009$.}
\label{fig:LT_h_dep}
\end{figure}

\begin{figure}[t!]
\centering
\subfigure{\includegraphics[scale=0.32,clip]{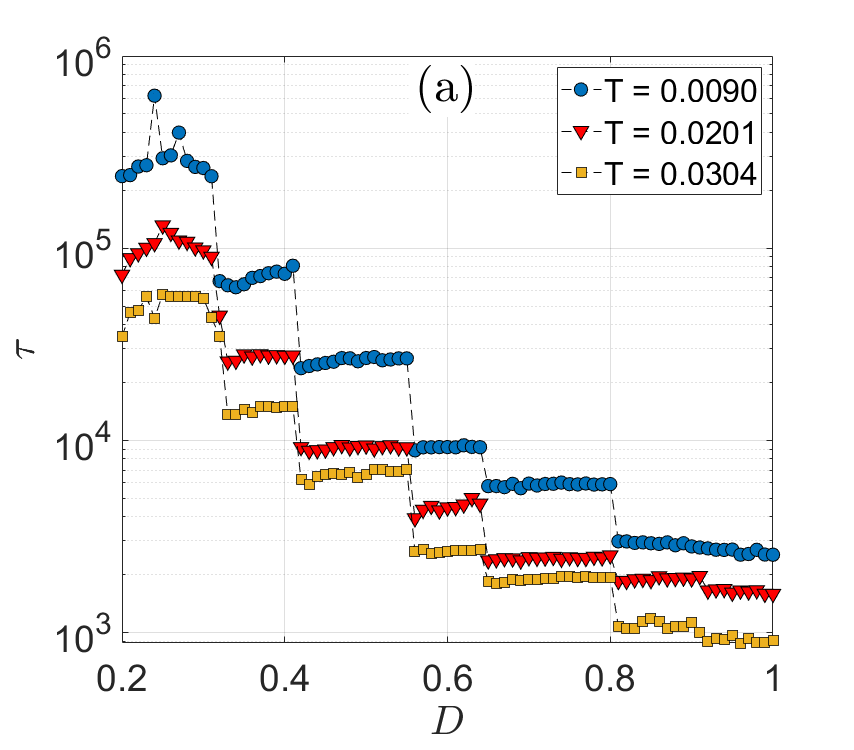}\label{fig:LT_D_dep}}
\subfigure{\includegraphics[scale=0.32,clip]{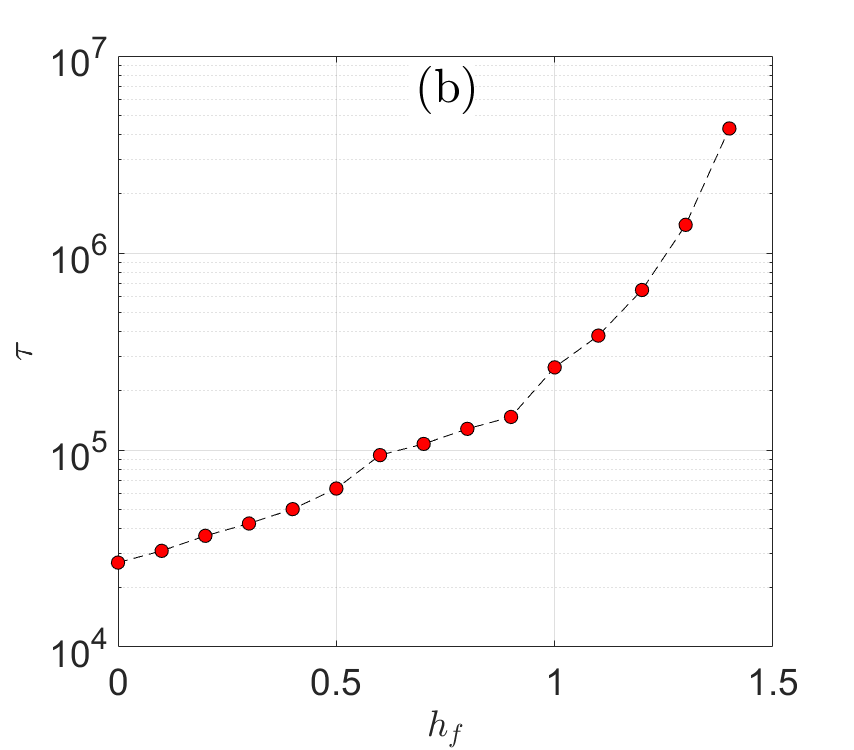} \label{fig:LT_HLT_dep}}\\
\subfigure{\includegraphics[scale=0.32,clip]{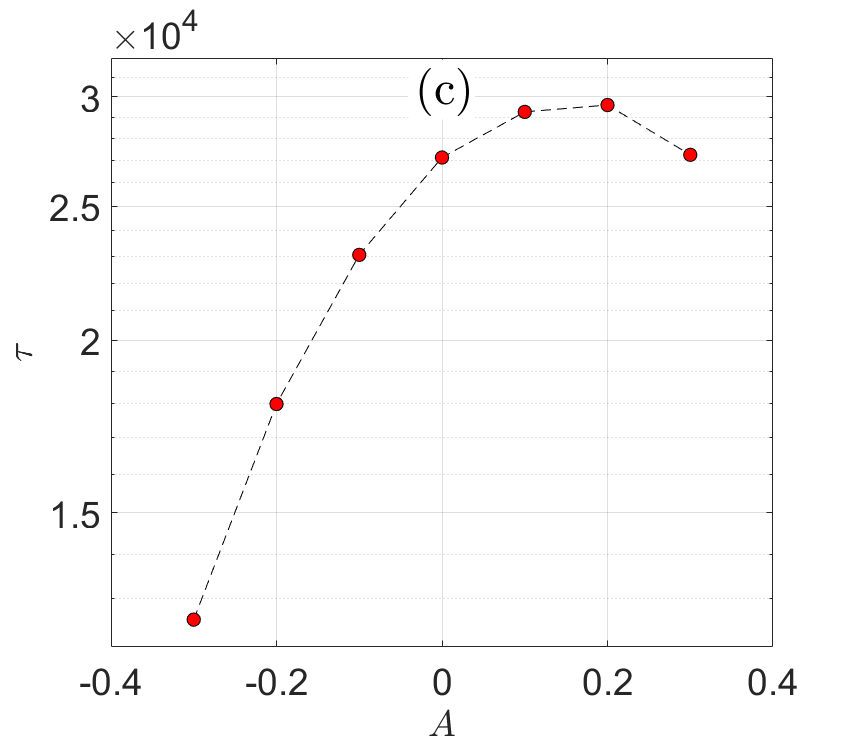}\label{fig:LT_A_dep}}
\caption{Skyrmion lifetime dependence on (a) the DMI strength $D$ for $A = 0$, $h = 3.2$, $h_f = 0$, and three different temperatures, (b) the final field $h_f$ for $D = 0.5$, $A = 0$, $h = 3.2$, $T = 0.009$, (c) the single-ion anisotropy $A$ for $D = 0.5$, $h = 3.2$, $h_f = 0$, $T = 0.009$.} 
\label{fig:LTdif}
\end{figure}

The effects of various parameters on the skyrmion persistence is demonstrated in Fig.~\ref{fig:LT_h_dep}. In the top row we show the skyrmion chirality evolution curves initialized from the SkX states obtained for $D = 0.5$ at different fields and $T=0.009$ (Fig.\ref{fig:LT_D_dep}) and $T=0.05$ (Fig.\ref{fig:LT_HLT_dep}). One can observe that the increasing field values, at which the initial configuration was generated, lowers the plateau height but it seems to have little to no influence on the skyrmion lifetime. The decrease of the plateau height results from the fact that the increasing field shrinks the skyrmions diameter by polarizing more spins further away from the core into its direction, thus making their arrangement on the lattice less closely packed. Consequently, upon quenching the field to zero the skyrmions relax to the state somewhat less regular and compact with smaller skyrmion chirality than at lower fields. On the other hand, as one would expect, with the increasing temperature thermal fluctuations progressively deform the skyrmions within the SkX state, which translates to the decreasing plateau height and eventually leads to their faster collapse.


The panels in the bottom row of Fig.~\ref{fig:LT_h_dep} show the skyrmion chirality evolution in the cases when the magnetic field is quenched from the initial $h=3.2$ to zero or small values $h_f$ (Fig.~\ref{fig:LT_H_pos}) and in the presence of the single-ion anisotropy (Fig.~\ref{fig:LT_with_ani_D050}). In the former case, in thermal equilibrium the fields $h_f$ correspond to the HL phase and, therefore, the skyrmion lifetime is finite. Nevertheless, under their influence it can be prolonged and the shape of the skyrmions within the plateau region better preserved. As evidenced in the behaviour of the curves presented in Fig.~\ref{fig:LT_with_ani_D050} the effect of the single-ion anisotropy is not as straightforward. On one hand, the increasing easy-axis anisotropy ($A<0$) appears to have an adverse effect on the skyrmion persistence. Namely, the plateau heights decrease and the curves tend to zero faster than for the isotropic case. On the other hand, the effect of the easy-plane anisotropy seems to depend on the anisotropy strength. Small enough $A>0$ tends to extend the skyrmion lifetime up to some value $A \approx 0.2$\footnote{In this case the chirality decay is slow and the skyrmion-like texture persists even beyond the inflextion point, as shown in the inserted snapshot.}, above which it starts to shrink.

Finally, we summarize the skyrmion lifetime dependence on various parameters in Fig.~\ref{fig:LTdif}. The dependence of $\tau$ on the temperature and the DMI presented in Fig.~\ref{fig:LT_D_dep} is quite apparent. Blue, red and yellow curves correspond to the temperatures $T = 0.009, 0.0201$ and $0.0304$. Evidently, even small increase in temperature shortens the skyrmions lifetime and even more so for lower DMI values. For a fixed temperature skyrmion configurations generated at higher DMI ($D = 1$) seem to be more fragile, with the difference in $\tau$ values reaching almost two orders of magnitude, compared to the smallest simulated value of $D = 0.2$. The fact that the lifetime changes in a step-wise fashion is most likely related with the discontinuous variation of the skyrmions' size with $D$~\cite{mohylna2021formation}. That would imply that bigger skyrmions formed at smaller DMI show better stability than smaller ones resulting from higher DMI. Fig.~\ref{fig:LT_HLT_dep} demonstrates the systematic increase of the lifetime if the magnetic field is quenched from some initial value within the SkX phase to some smaller value $h_f$, corresponding to the HL phase. On the other hand, Fig.~\ref{fig:LT_A_dep} shows that the effect of the single-ion anisotropy on the skyrmion lifetime is not as dramatic as in the cases of the above parameters. The maximum value is achieved at relatively small values of the easy-plane anisotropy, while larger values of both the easy-plane and easy-axis anisotropy result in shortening of the skyrmion lifetime.


\section{Summary}
We have studied the frustrated antiferromagnetic Heisenberg model on the triangular lattice with the Dzyaloshinskii-Moriya interaction (DMI) in the presence of the external magnetic field and the single-ion anisotropy. We applied the parallel tempering MC simulations to study the phase diagram evolution with the changing anisotropy in the regimes of a moderate and strong DMI, and the standard MC with the Metropolis dynamics to probe the persistence of the skyrmion lattice (SkX) phase upon quenching the field to zero or finite small values.

Our results suggest that a strong single-ion anisotropy has overall destructive effects on the SkX phase. While the easy-axis type encourages formation of the UUD phase within SkX and with further increase its gradual growth primarily at the cost of SkX, the easy-plane type tends to extend the HL phase and thus to reduce the area of SkX from the low fields side. This effect becomes less pronounced in the case of a stronger DMI since SkX becomes more robust and much higher anisotropy is required to produce a noticeable change in the phase diagram topology. On the other hand, a small anisotropy can be beneficial for the SkX phase. In the systems with a moderate DMI, a small easy-plane anisotropy extends the field range of SkX towards lower fields, albeit it slightly reduces its temperature range. For the systems showing larger DMI, a small easy-axis anisotropy can be beneficial by shifting the field window of the SkX appearance to lower values, while preserving or even slightly extending its temperature window.

As for the persistence of the skyrmions in the absence of the external magnetic field, the most favorable conditions correspond to low temperatures, small DMI and small easy-plane anisotropy values. Nevertheless, there are dramatic differences between the effects of the individual parameters. While the effect of the single-ion anisotropy is rather marginal, by decreasing temperature or DMI the skyrmion lifetime can be increased over several orders of magnitude. In the latter case the increase occurs in a step-wise fashion, which is related to the discrete increase (decrease) of the skyrmion number (size) due to the lattice finiteness, as shown in our previous study~\cite{mohylna2021formation}. In the same study we estimated the lower DMI threshold for the existence of SkX as $D_t \approx 0.02$ and, thus, by extrapolating one can expect a remarkable SkX phase stability for DMI approaching this value. Finally, the skyrmion lifetime can also be prolonged if the external field in quenched not to zero but some small finite values, corresponding to the stable HL phase.

We believe that our findings can be relevant and useful in synthesizing various experimental realizations of AFM skyrmion lattice using materials engineering approach. Recently, possibilities of integrating transition metal dichalcogenides with magnetic transition metals to realize AFM triangular lattices capable of hosting such a skyrmion lattice have been explored and several potential candidates, such as Cr/MoS\textsubscript{2}, Fe/MoS\textsubscript{2}, and Fe/WSe\textsubscript{2}, have been proposed~\cite{fang2021}. Knowledge of effects of different parameters on the skyrmion lattice existence and stability are essential in their fine-tuning to stabilize the skyrmion phase under experimentally feasible or technologically favourable conditions. 

\section*{Acknowledgment}
This work was supported by the Scientific Grant Agency of Ministry of Education of Slovak Republic (Grant No. 1/0531/19) and the Slovak Research and Development Agency (Contract No. APVV-16-0186). Part of computations was held on the basis of the HybriLIT heterogeneous computing platform (LIT, JINR)~\cite{adam2018ecosystem}. 


\end{document}